\documentclass[12pt]{spieman}  % 12pt font required by SPIE;
\usepackage{amsmath,amsfonts,amssymb}
\usepackage{graphicx}
\usepackage{setspace}
\usepackage{tocloft}
\usepackage{lineno}
\usepackage{subcaption}
\linenumbers
\title{Single-shot, spatially resolved spectropolarimetry with stress engineered optics}

\author[a*]{David Spiecker}
\author[a]{Thomas Brown}

\affil[a]{Institute of Optics, University of Rochester, 480 Intercampus Dr, Rochester, NY 14627, USA\\}

\cftpagenumbersoff{figure}
\cftpagenumbersoff{table} 
\begin{document} 
\nolinenumbers
\maketitle

\begin{abstract}
We describe, and test, a method of obtaining spectrally and spatially sampled Stokes polarimetry measurements based on star test polarimetry enabled by a stress engineered optic (SEO).  When the SEO is placed in the pupil plane of an imaging system, the space-variant birefringence of the SEO combined with the intrinsic wavelength dependence of the retardance allows for point spread functions from distinct laser wavelengths to be both spectrally and polarimetrically distinguished in a single frame measurement designed to enable single shot spectropolarimetry of individual laser pulses. In the experiment, the beam is sampled using a lenslet array typical of that used in a Shack-Hartmann sensor with the PSF of each spot relayed to a sensor. Experiments show angular errors as small as 100 mRad on the Poincar\'e sphere, with red and blue measurements outperforming green. Potential applications include polarization sampling of extreme lasers as well as spectropolarimetry of display devices based on liquid crystals.
\end{abstract}

% Include a list of up to six keywords after the abstract

% Include email contact information for corresponding author
{\noindent \footnotesize\textbf{*}David Spiecker,  \linkable{davidspiecker@rochester.edu} }

\begin{spacing}{2}   % use double spacing for rest of manuscript

\section{Introduction}
\label{sect:intro}
Spectropolarimetry is the acquisition of polarimetric information encoded into some spectral content which may also vary over a scene of interest. To appreciate some of the challenges in spectropolarimetry, an interpretation of the data acquired in spectropolarimetry as discussed by Sabatke\cite{sabatke_snapshot_2002} is shared here. "The data a spectropolarimeter acquires can be interpreted as an image of a 4-D volume, since a measure of radiance is obtained for four independent variables or indices: two spatial variables ~($x,y$), wave number ($\sigma$), and the Stokes vector index ($j$). Note, however, that the Stokes vector index has only four possible values (the integers from 0 to 3), whereas the $x$, $y$, and $\sigma$ dimensions are each segmented into a greater number of intervals. We refer to this 4-D volume as the spectropolarimetric hypercube." Performing a 1-, 2-, or 3-D imaging of this data volume requires scanning through the remaining dimensions. Channeled spectropolarimeters image this data volume along the $\sigma$ and $j$ dimensions by encoding this information in a modulated intensity pattern along two orthogonal spatial dimensions on a sensor\cite{todorov_spectrophotopolarimeter_1992, oka_spectroscopic_1999, sabatke_snapshot_2002, knitter_spectrally_2011}. Different spectropolarimetric methods make trade-offs in measurement between spatial and spectral dimensions with differing temporal requirements and are reviewed by Tyo\cite{tyo_review_2006}. Recent years have also seen an interest in metasurface design for single frame/single shot full Stokes imaging polarimetry\cite{rubin_matrix_2019,rubin_imaging_2022,li_metasurface-enabled_2025}; however, single frame/single shot measurements that combine spatial, spectral, and full Stokes information remain a challenge.  

In this paper, we describe a spectropolarimetric method that has the capability of sampling all four dimensions of the spectropolarimetric hypercube in a single irradiance measurement. Our method is a spatially resolved spectropolarimeter using stress-engineered optics (SEOs) that can acquire spectropolarimetric information in a 2-D scene of interest in a single frame and thus is capable of single-shot measurements. In essence, we decrease the sampling resolution along the $x$, $y$, and $\sigma$ dimensions of the spectropolarimetric hypercube, which allows us to perform spatial binning of spectropolarimetric information. Within a single spatial bin, the spectrum is likewise binned so that polarimetric information can be extracted from each spectral bin through a single irradiance pattern. We consider a beam that comprises multiple, discrete wavelengths in such a way that each spectral component has polarization properties (Stokes parameters) that are uncorrelated with one another and that a composite point spread function (PSF) can be expressed as a linear superposition of the PSFs for each spectral component. This superposition means that we can take what is known about star-test polarimetry with a monochromatic input and treat polychromatic inputs as a superposition of the system's response for each input wavelength. The key optical element in the system is an SEO, that was applied to image-sampling Stokes polarimetry in prior work by Zimmerman and Brown \cite{zimmerman_star_2016} and to sampling the Stokes parameters in a multicore fiber by Sivankutty et al.\cite{sivankutty_single-shot_2016}. In these systems, an SEO is most easily implemented as plane-parallel window with an external, peripheral stress distribution that in the simplest cases, has trigonal symmetry.
\begin{figure}
    \centering
    \includegraphics[width=0.85\linewidth]{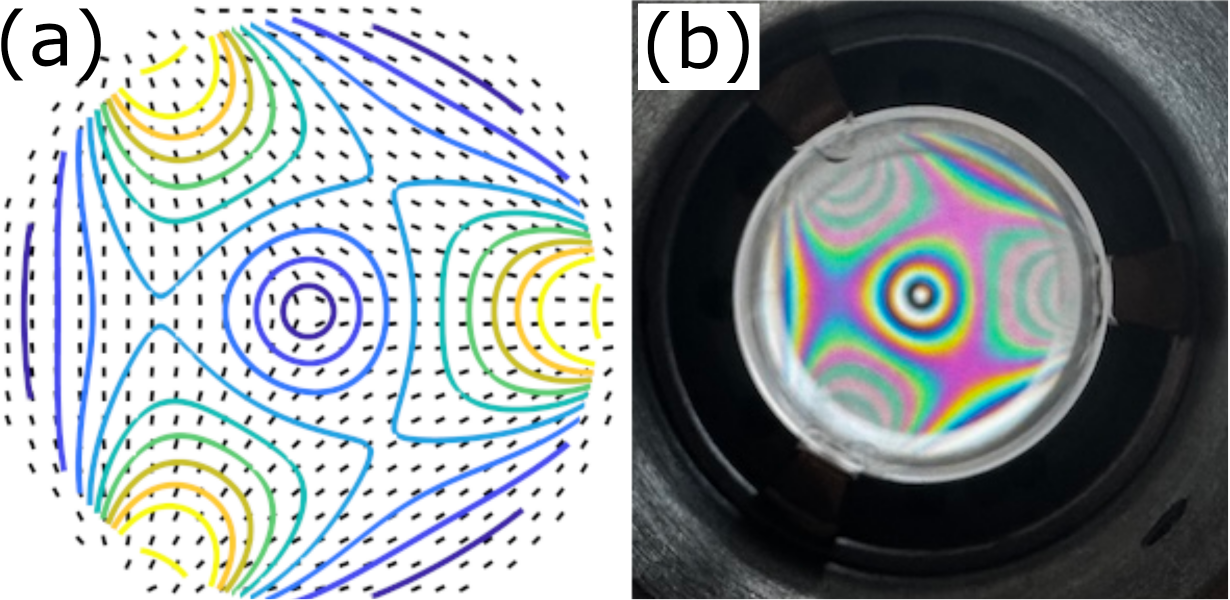}
    \caption{(a) Theoretical model of an SEO, where the short, black lines indicates the slow axis orientation and the colored lines indicate contours of equal retardance in an SEO. (b) A physical SEO viewed with a right-circular polarized input and a matching analyzer, demonstrating the phase retardance patterns present in SEOs. The phase retardance pattern presents itself through the observed intensity of light and the phase retardance in the central region varies linearly.}
    \label{fig:theory_exp_model_SEO}
\end{figure}

\subsection{Stress engineered optics}
The behavior of stress in an SEO is well described using the theoretical model developed by Yiannopoulos \cite{yiannopoulos_general_1999} and applied to SEOs by Brown and Beckley \cite{brown_stress_2013}. With the knowledge of the stress distribution in an optical window, the slow axis orientation and the phase retardance at every point in the window can be predicted. Figure \ref{fig:theory_exp_model_SEO} shows a typical result of such a calculation assuming threefold symmetry with external stress distributed tightly over three regions separated by 180$^\circ$. A key parameter of SEOs is the dimensionless stress parameter $c$ which describes the rate of change of the phase retardance over normalized radial distances in the central region of SEOs. The dimensionless stress parameter is defined through the phase retardance in an optical element, which is defined as 
\begin{equation}
    \delta(\rho,\phi)= \frac{ 2\pi}{\lambda}t\Gamma(\rho,\phi),
    \label{eq:phase_retardance}
\end{equation}
where $\rho$ is the normalized (to the radius of the optical window) radial direction and $\phi$ is the azimuthal direction in a cylindrical polar coordinate system, $t$ is the thickness of the SEO in the direction of the propagation of light, $\lambda$ is the wavelength of the light, and $\Gamma$ is the stress birefringence of the medium. The stress birefringence depends on the loading geometry of the optical window and the window's material properties. Of particular interest is the region near the center of the SEO (up to approximately $\rho=0.2$). In this region, the phase retardance varies only in the radial direction and the fast axis orientation varies only in the azimuthal direction. For SEOs with trigonal loading geometry such as the one used in our experiments, the phase retardance varies linearly and can be expressed as 
\begin{equation}
    \delta(\rho)=c\rho,
    \label{eq:retardance_power_law}
\end{equation}
where $c$ is the dimensionless stress parameter which is proportional to the external applied force and represents the rate of change of the phase retardance over normalized radial distances\cite{brown_stress_2013}. By inspection of Eqs. \ref{eq:phase_retardance} and \ref{eq:retardance_power_law}, it can be understood that for any given amount of externally applied force, the dimensionless stress parameter $c$ is inversely proportional to the wavelength of light producing wavelength-dependent rates of phase retardance over radial distances. 

\begin{figure}[h]
	\includegraphics[width=0.8\linewidth]{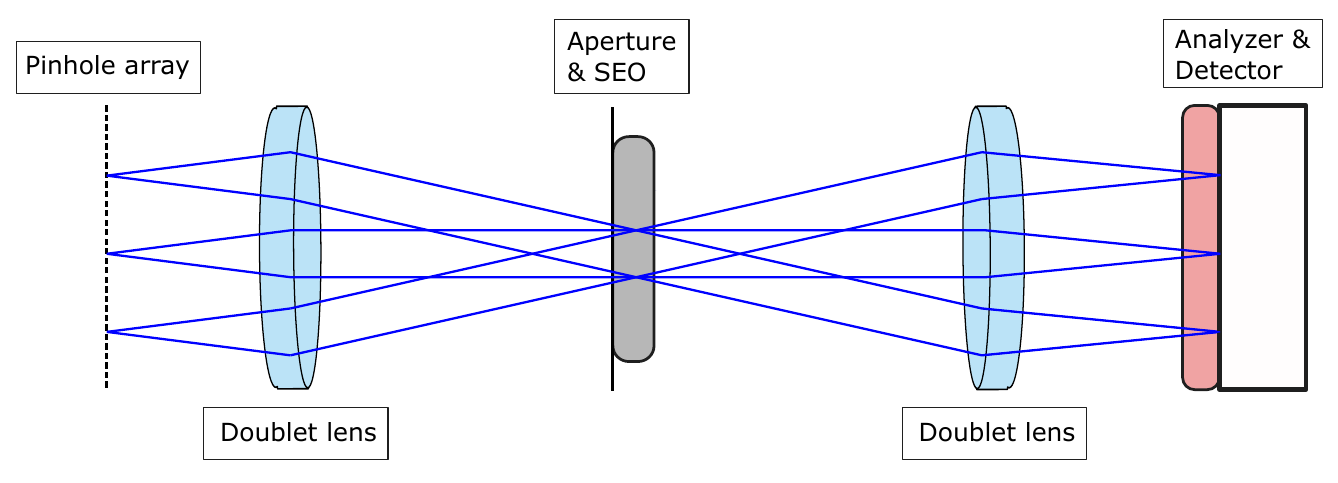} 
	\centering
	\caption{Star Test Image Sampling Polarimeter as developed by Zimmerman and Brown in which the pinhole array creates an array of point sources sampling a scene of interest\cite{zimmerman_star_2016}.}
	\label{fig:STIP}
\end{figure}

\begin{figure}[h]
    \centering
    \includegraphics[width=0.8\linewidth]{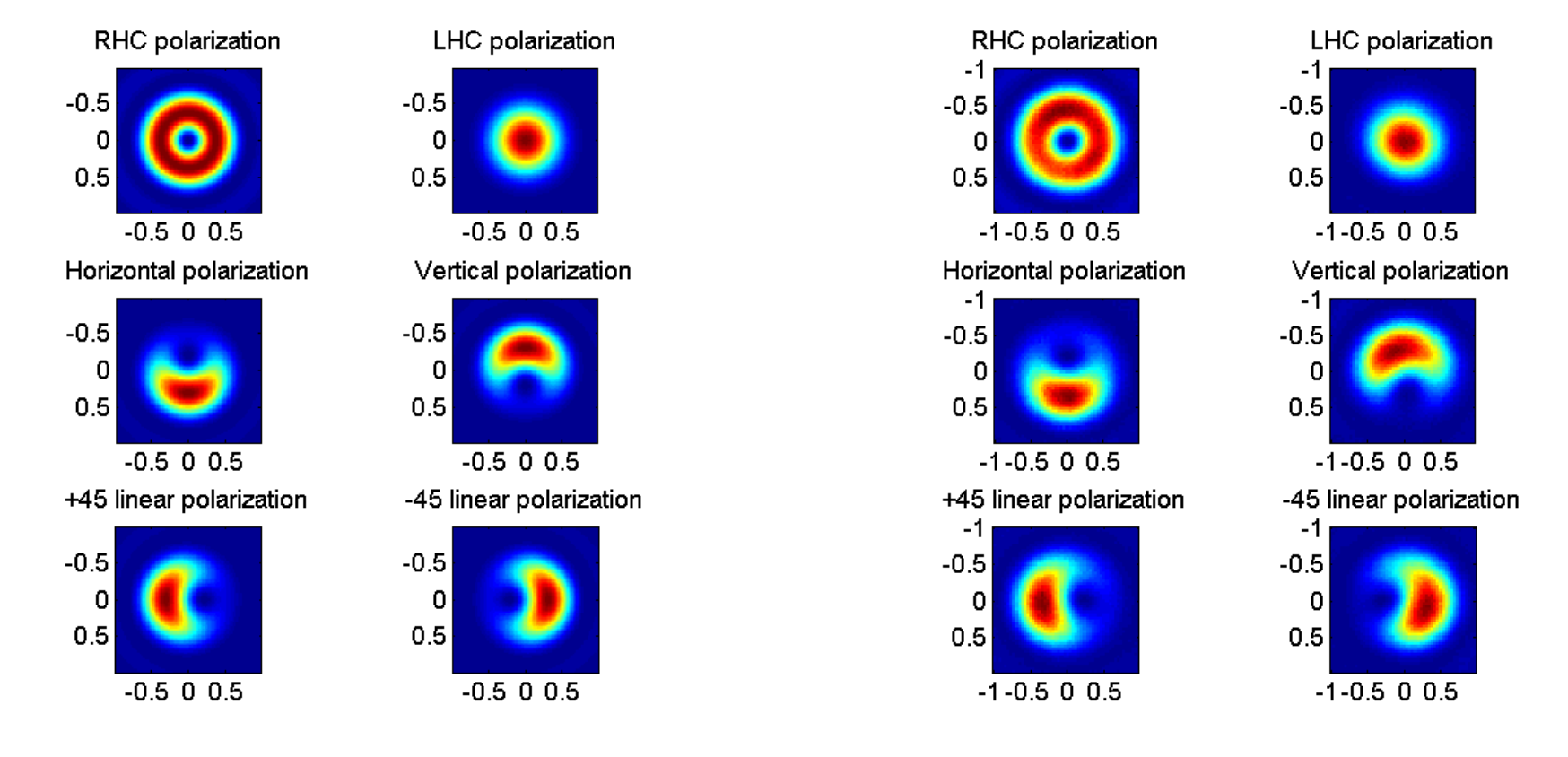}
    \caption{PSFs of reference polarization states as predicted (left) and measured (right) by Ramkhalawon. (Reproduced with permission from Ramkhalawon, Brown, and Alonso\cite{ramkhalawon_star_2012}.)}
    \label{fig:SEO_PSFs}
\end{figure}

When the SEO is placed in the Fourier plane of a 4F imaging system (Fig. \ref{fig:STIP}) and a point source is imaged, the resulting point spread function (PSF) (with a polarization analyzer) acquires an irradiance pattern that is unique to the input polarization state as demonstrated in Fig. \ref{fig:SEO_PSFs} \cite{ramkhalawon_star_2012,ramkhalawon_imaging_2013}.

\begin{figure}[ht]
    \centering
    \includegraphics[width=1\linewidth]{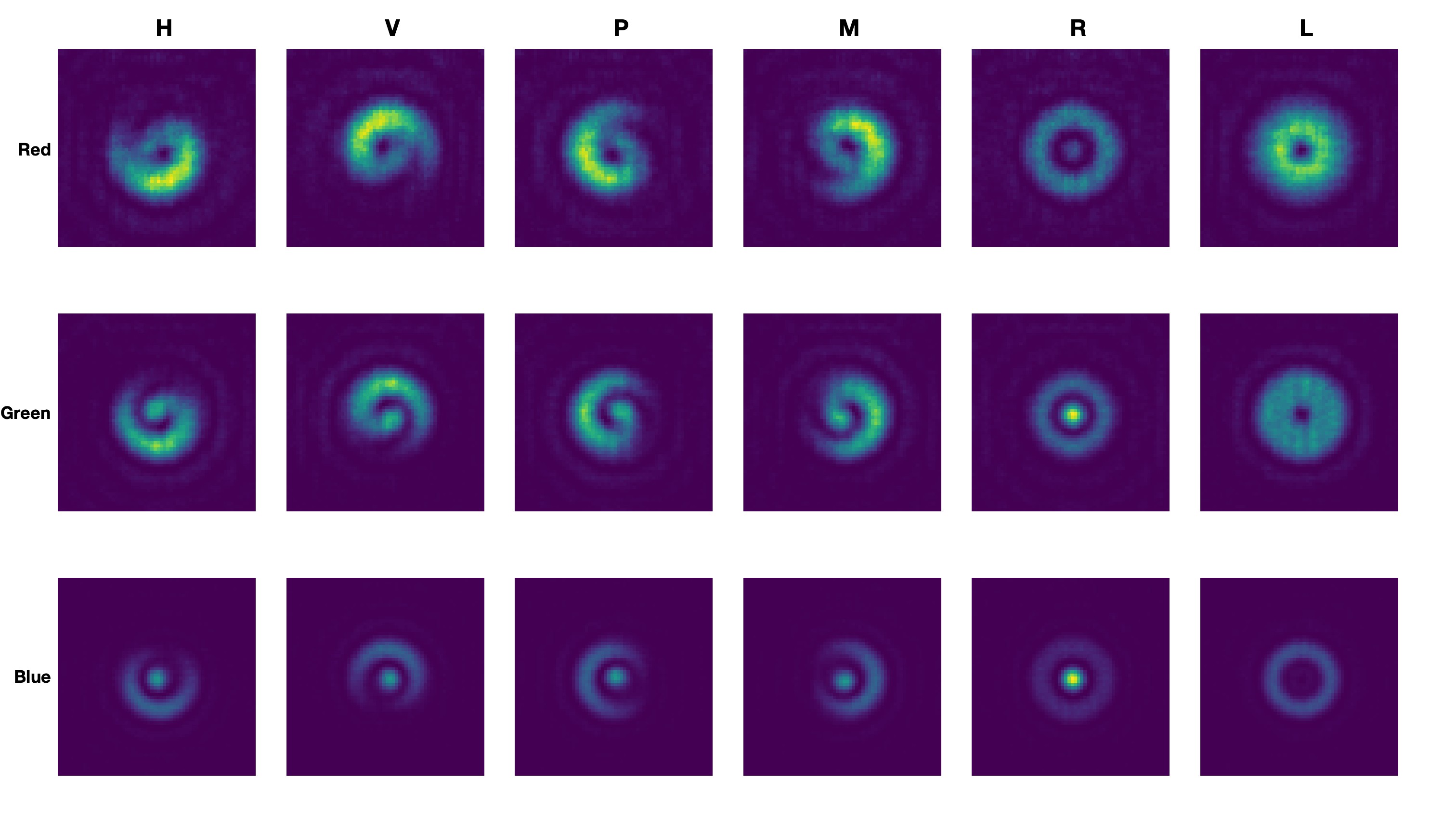}
    \caption{Simulated PSFs of the six reference polarization states for each of the three input wavelengths at $c_{ref}=2.88\pi$}
    \label{fig:poly_cal_psfs}
\end{figure}

Due to the wavelength dependence of the dimensionless stress parameter $c$, each input wavelength has an effective value of $c$ for a given amount of pressure on the SEO. For clarity in discussing values of $c$, this paper focuses on three input wavelengths: 635, 520, and 405 nm, each labeled as "red", "green", and "blue" or "$\lambda_{635}$", "$\lambda_{520}$", and "$\lambda_{405}$" respectively. Likewise, due to the wavelength dependence of $c$, 520 nm is used as a reference wavelength when discussing values of $c$ for the SEO and is indicated with $c_{ref}$. Input wavelengths with the same polarization state will produce an unique irradiance pattern relative to each other as a result of the wavelength dependence of $c$. This is demonstrated with simulated PSFs for input wavelengths of 630 nm, 520 nm, and 405 nm as shown in Fig. \ref{fig:poly_cal_psfs}.

The polarization state of an input field are represented by the Stokes parameters $S_0, S_1, S_2,$ and $S_3$. The polarization states described by the Stokes parameters $S_1, S_2,$ and $S_3$ are referenced as H, V, P, M, R, L for horizontal, vertical, +45, -45, right-circular, and left-circular respectively. Since the irradiance pattern at the sensor is uniquely related to the input polarization, the system can be modeled with a measurement matrix $\mathbf{M}$ as shown in Eq. (\ref{eq:measurement_matrix_equation}),
\begin{equation}
    I(\vec{x})=\mathbf{M}(\vec{x})\vec{S},
    \label{eq:measurement_matrix_equation}
\end{equation}
where $I$ is the irradiance pattern, $\vec{x}$ is a vector of all image points, and $\vec{S}$ is the Stokes vector representing the input polarization state. The measurement matrix (a $j\times 4$ matrix where $j$ is the number of elements in $\vec{x}$) can be constructed from the product of the pseudoinverse of the matrix of N concatenated Stokes vectors (represented as a $4 \times N$ matrix with each column representing an input polarization state) and their corresponding irradiance patterns (represented as a $j \times N$ matrix) as shown in Eq. (\ref{eq:measurement_matrix_construction}).

\begin{equation}
    \mathbf{M}(\vec{x})=[I_{\vec{x},1}, \; I_{\vec{x},2}, \; \cdots , \; I_{\vec{x},N}] 
    \begin{bmatrix}
        S_{0,1}, \; S_{0,2}, \; \cdots , \; S_{0,N} \\
        S_{1,1}, \; S_{1,2}, \; \cdots , \; S_{1,N} \\
        S_{2,1}, \; S_{2,2}, \; \cdots , \; S_{2,N} \\
        S_{3,1}, \; S_{3,2}, \; \cdots , \; S_{3,N} \\
    \end{bmatrix}^{-1}
    \label{eq:measurement_matrix_construction}
\end{equation}
This measurement matrix is constructed with a monochromatic input. For polychromatic inputs, the input can be understood as a linear superposition of multiple, discrete wavelengths. This superposition means that we express the unknown irradiance pattern as a product of the wavelength-dependent measurement matrix and the wavelength-dependent Stokes vector. This can be expressed as
\begin{equation}
    I(\vec{x})=\mathbf{M}_{\lambda}(\vec{x})\vec{S}_{\lambda},
    \label{eq:poly_STIP}
\end{equation}
where $\mathbf{M}_\lambda(\vec{x})$ is the $j \times 4n$ (with $j$ being the number of elements in $\vec{x}$ and $n$ being the number of wavelengths) concatenated monochromatic measurement matrices of each wavelength, expressed as 
\begin{equation}
    \mathbf{M}_{\lambda}(\vec{x})=[\mathbf{M}_{\lambda_1}(\vec{x}), \mathbf{M}_{\lambda_2}(\vec{x}),\;...\;,\mathbf{M}_{\lambda_n}(\vec{x})], 
    \end{equation}
    and $\vec{S}_\lambda$ is the $4n \times 1$ column vector of concatenated Stokes vectors of each wavelength, expressed as
\begin{equation}
    \vec{S}_{\lambda}=[\vec{S}_{\lambda_1},\vec{S}_{\lambda_2}, \;...\; ,\vec{S}_{\lambda_n}].
\end{equation}
The unknown polychromatic Stokes vector $\vec{S}_{U,\lambda}$ may be retrieved from the measurement of a polychromatic irradiance pattern if $\mathbf{M}_\lambda(\vec{x})$ is known. By performing a pseudoinverse operation on the wavelength-dependent measurement matrix to obtain a data reduction matrix $\mathbf{M}_{\lambda}(\vec{x})^{-1}$, the polarization state of each wavelength in the input can be retrieved from the product of the data reduction matrix and the measured irradiance pattern $I_{meas}(\vec{x})$ as expressed by Eq. (\ref{eq:poly_Stokes_retrieval}).
\begin{equation}
   \vec{S}_{U,\lambda}=\mathbf{M}_{\lambda}(\vec{x})^{-1} I_{meas}(\vec{x})
    \label{eq:poly_Stokes_retrieval}
\end{equation}

\begin{figure}
    \centering
    \includegraphics[width=0.5\linewidth]{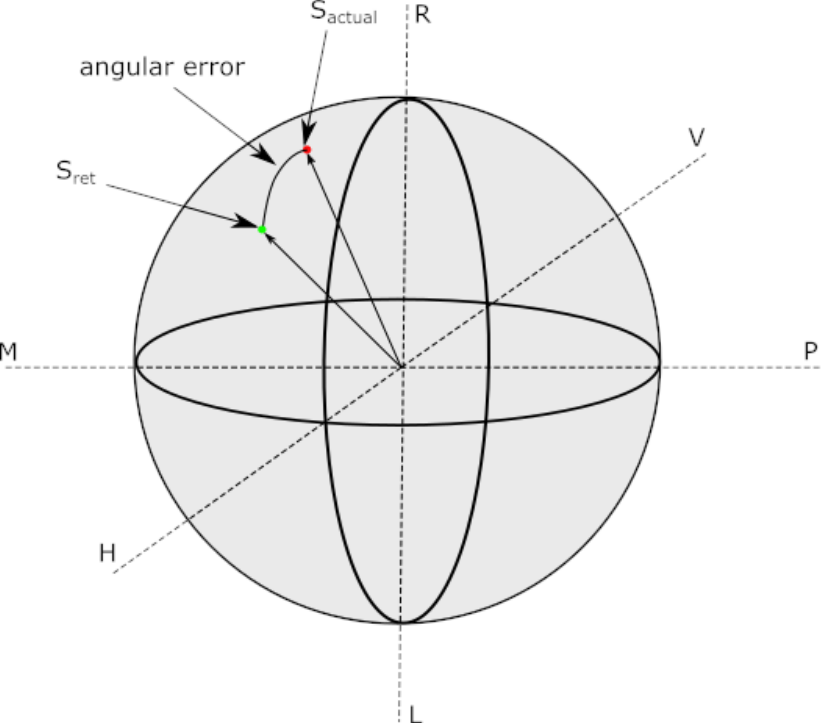}
    \caption{Angular error as illustrated on the Poincaré sphere}
    \label{fig:angular_error}
\end{figure}

The performance of the system is determined by the polarimetric angular error. The polarimetric angular error is defined as the angular separation of the input's Stokes vector and the retrieved Stokes vector as represented on the Poincaré sphere (Fig. \ref{fig:angular_error}).
Previous work in simulating spectropolarimetry with SEOs have shown that the angular error in the polychromatic Stokes vector retrieval of a single point source is dependent on values of $c_{ref}$\cite{spiecker_single-shot_2024}. The angular error for each input wavelength varies based on the value of $c_{ref}$ and is explored further with an array of point sources for three different values of $c_{ref}$ in this paper. 

\section{Materials and methods}
\begin{figure}
    \centering
    \includegraphics[width=0.75\linewidth]{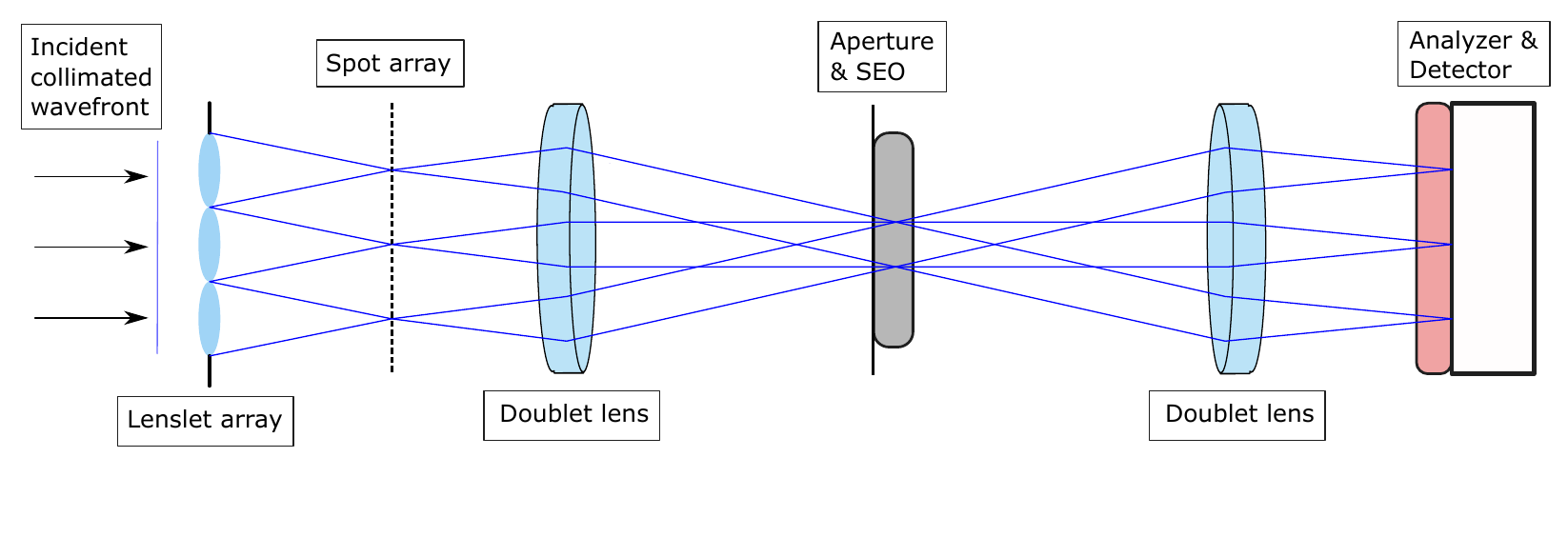}
    \caption{Schematic of the experimental setup for single-shot, spatially resolved spectropolarimetry}
    \label{fig:spectropolarimetry_schematic}
\end{figure}

The experimental arrangement is shown in Fig. \ref{fig:spectropolarimetry_schematic}. A mounted fused silica lenslet array (ThorLabs MLA150-7AR-M) was used to create an array of point sources. A pair of 25.4 mm diameter achromatic doublets with a focal length of 125 mm were used to create the 4F imaging system. A hydraulic pressure SEO\cite{spiecker_stress_2025} and a circular aperture with a radius that is 20\% the SEO's radius (2.54 mm) was placed at the Fourier plane of the 4F system. A CMOS image sensor (Basler acA5472-17um) with a right-circular analyzer in front of it was placed at the image plane of the 4F system. Three fiber-coupled benchtop lasers at 405, 520, and 630 nm each with air-spaced doublet fiber collimators for each wavelength are used as an input source. The polarization state of the input is prepared with a linear polarizer, an achromatic quarter-wave plate (ThorLabs AQWP10M-580), and an achromatic half-wave plate (ThorLabs AHWP10M-580). The prepared polarization state is measured with a polarimeter (ThorLabs PAX1000VIS) as the input beam travels to the lenslet array. 

Since each point source produced by the lenslet array has its own PSF, each lenslet requires its own measurement matrix to be constructed. The measurement matrix is constructed in accordance to Eq. (\ref{eq:measurement_matrix_construction}), using six reference states for each wavelength. Three additional states are recorded for each wavelength but are not used in the construction of the measurement matrix. For each wavelength, a set of five states is defined — two that were used in constructing the measurement matrix and three that were not. Because a polychromatic input is understood to be the superposition of each individual input wavelength, additional polychromatic input states are generated by selecting one state from each wavelength’s set and combining them. Doing this creates a larger data set, enabling meaningful statistical evaluation of the polychromatic Stokes vector retrieval from a limited set of input polarization states for each wavelength. Since each PSF has its own measurement matrix, the angular error of the polychromatic Stokes vector retrieval of each PSF is determined and used in the statistical analysis. 

\section{Results}
\begin{figure}
    \centering
    \includegraphics[width=0.75\linewidth]{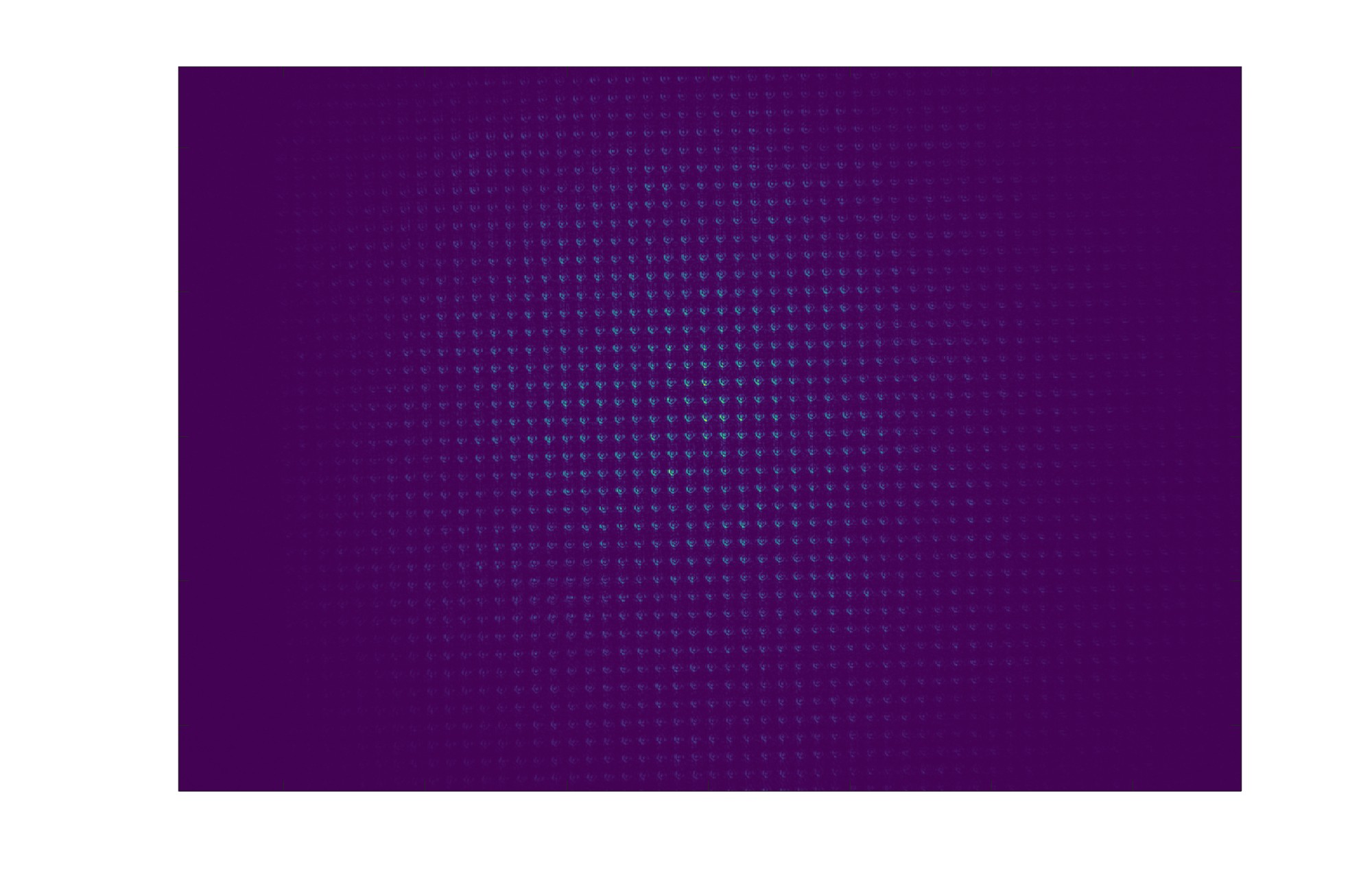}
    \caption{Measured irradiance pattern of H-polarized $\lambda_{405}$ input with an observable irradiance envelope.}
    \label{fig:detected_irradiance}
\end{figure}

\begin{figure}
    \centering
    \includegraphics[width=0.85\linewidth]{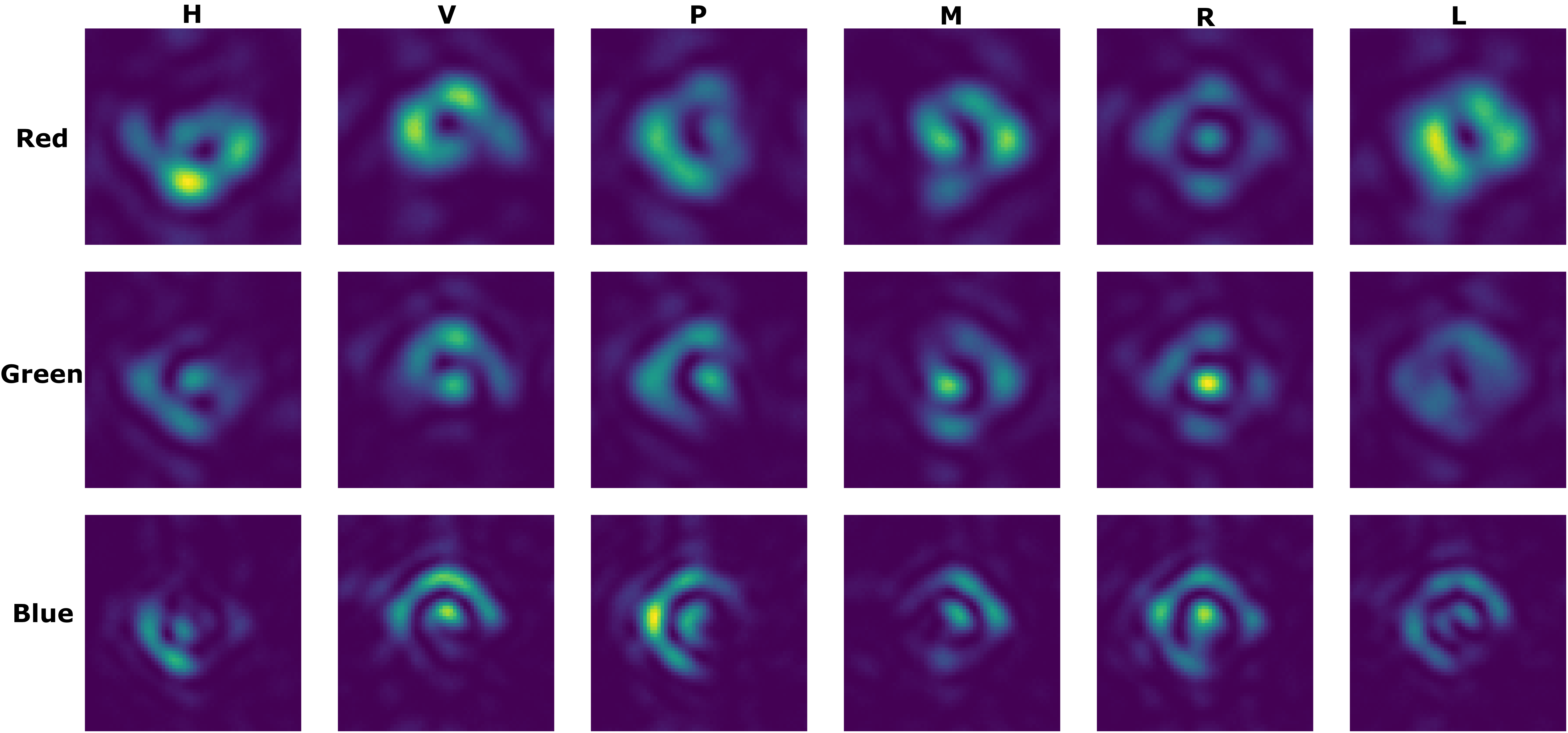}
    \caption{Experimental PSFs of the six reference polarization states for each of the three input wavelengths produced by a single lenslet at $c_{ref}=2.88\pi$}
    \label{fig:sample_psfs}
\end{figure}

All three input wavelengths, $\lambda_{635}$, $\lambda_{520}$, and $\lambda_{405}$, under-filled the lenslet array resulting in an array of PSFs with intensities corresponding to the incident irradiance envelope. A measured irradiance pattern is shown as an example in Fig. \ref{fig:detected_irradiance}. Usable PSFs and their locations are identified by setting a minimum degree of correlation using a normalized cross correlation with a Gaussian irradiance pattern and the irradiance pattern with R-polarized input (which contains an irradiance pattern resembling a Gaussian irradiance pattern). 
The irradiance pattern produced by a single lenslet for all input wavelengths and reference polarization states are shown in Fig. \ref{fig:sample_psfs}.
Deviations from the ideal PSFs for reference states occur from aberrations and other imperfections in the optical system. Those effects remain constant over time and are consistent for each input polarization state. With the construction of the measurement matrix for each PSF, those effects are considered a part of the system response to an input. The recorded reference PSFs still exhibit irradiance patterns that match expectations and have the appropriate relationship to each other. 

\begin{figure}
    \centering
    \includegraphics[width=.60\linewidth]{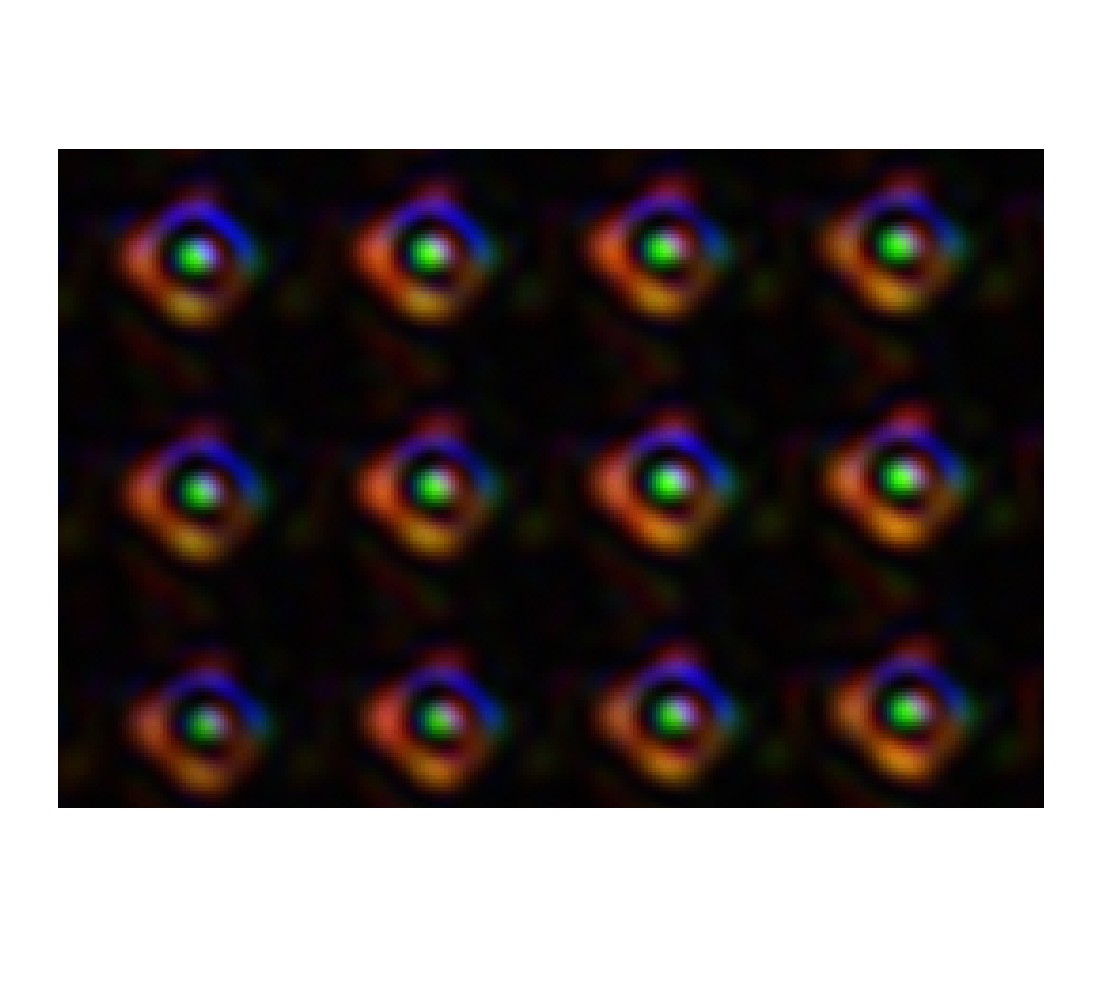}
    \caption{Visualization of a sample polychromatic input at $c_{ref} = 2.88\pi$ with the combination of the following Stokes vectors: Red Stokes vector: [1.000, 0.228, 0.622, 0.657], Green Stokes vector: [1.000, 0.708, -0.165, 0.657], and Blue Stokes vector: [1.000, -0.258, 0.602, 0.609] using experimental data. The image is magnified such that the polychromatic irradiance patterns are clearly visible.}
    \label{fig:polychromatic_input}
\end{figure}

\begin{figure}
    \centering
    \includegraphics[width=\linewidth]{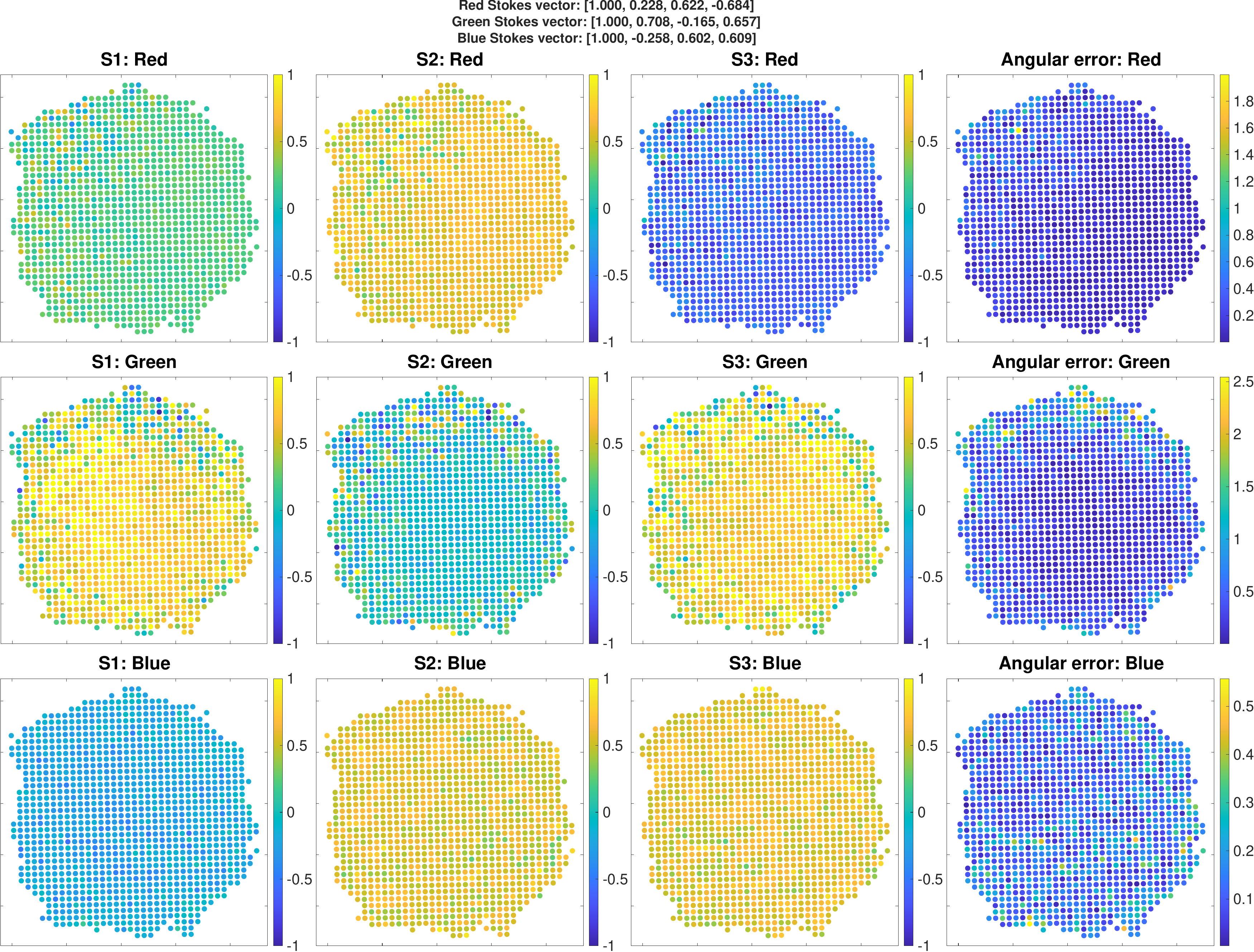}
    \caption{Spatially resolved Stokes parameter retrieval and angular error (measured in rad) at $c_{ref} = 2.88\pi$ of the same irradiance pattern shown in Fig. \ref{fig:polychromatic_input}.}
    \label{fig:spatially_resolved_angular_error}
\end{figure}

\begin{figure}
    \centering
    \includegraphics[width=\linewidth]{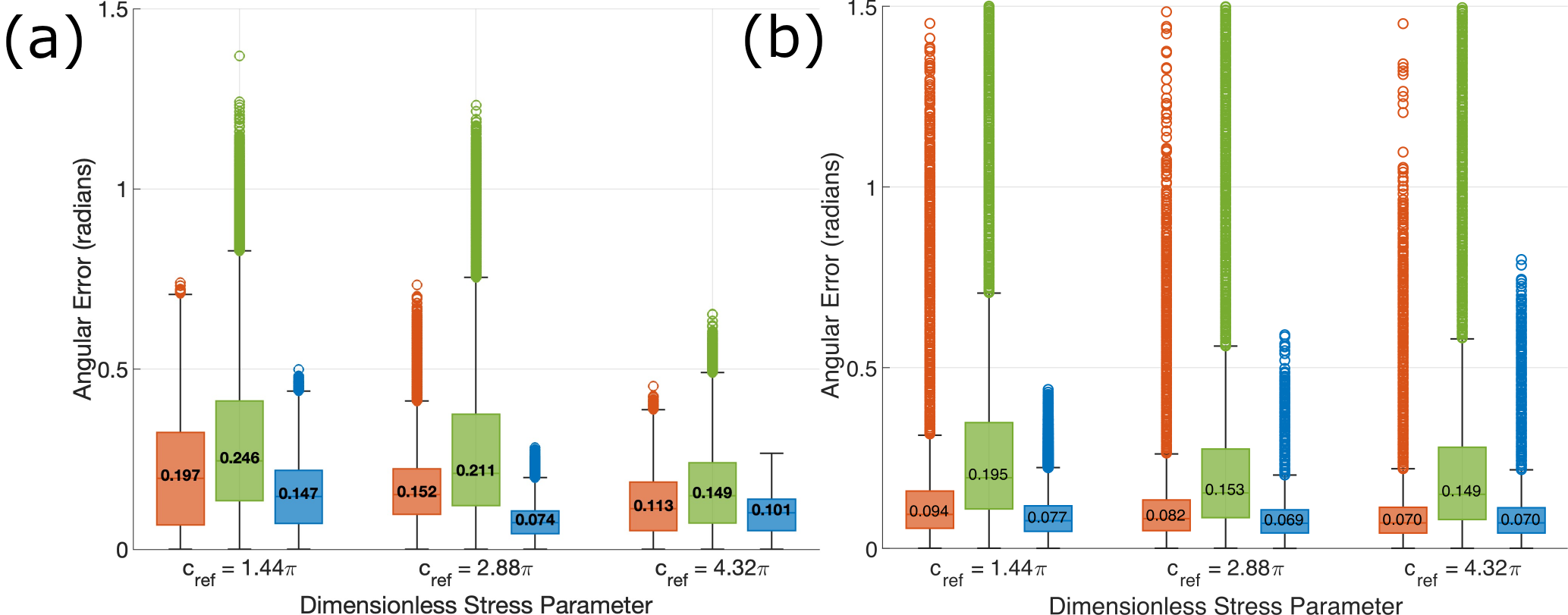}
    \caption{Box plot of angular error in polychromatic Stokes vector retrieval. The number in the middle of the box indicates the median value of the angular error. Each box's color corresponds to their respective input wavelengths. (a) Simulated polychromatic Stokes vector retrieval. (b) Experimental polychromatic Stokes vector retrieval.}
    \label{fig:angular_error_box_plot}
\end{figure}

\begin{figure}
    \centering
    \includegraphics[width=\linewidth]{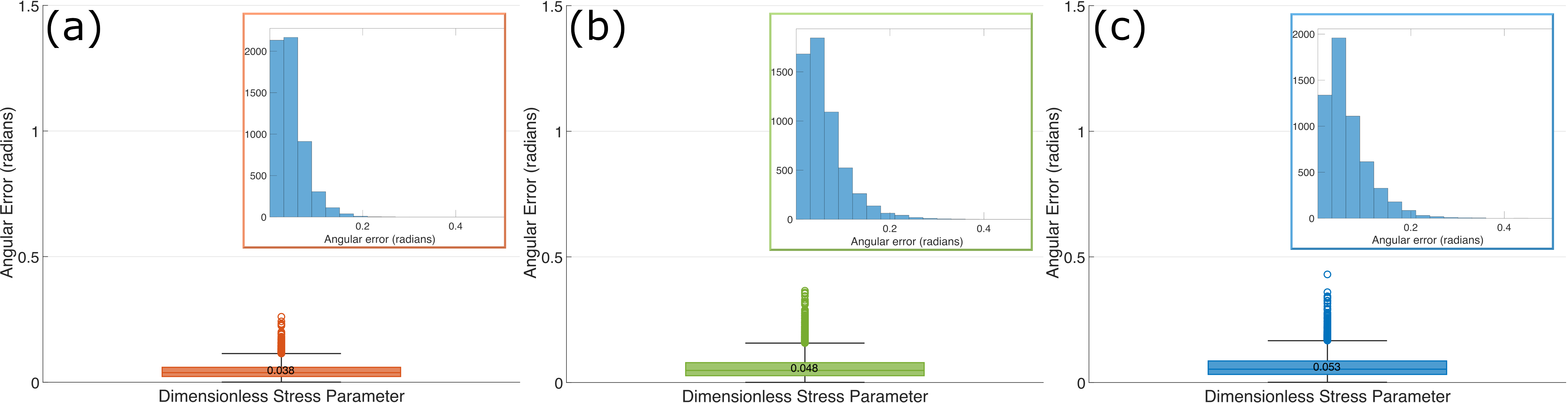}
    \caption{Box plot of angular error in monochromatic Stokes vector retrieval for $c_{ref}=2.88\pi$. The number in the middle of the box indicates the median value of the angular error. Each box plot contains an inset displaying the histogram of the plotted data for their respective input wavelength. Each box's color corresponds to their respective input wavelengths. (a) Red input with a median value of 0.038 rad. (b) Green input with a median value of 0.048 rad. (c) Blue input with a median value of 0.053 rad.}
    \label{fig:monochromatic_angular_error}
\end{figure}

One such polychromatic input is shown in Fig. \ref{fig:polychromatic_input} and the corresponding analysis of the Stokes vector retrieval is shown in Fig. \ref{fig:spatially_resolved_angular_error}.
The spatially resolved Stokes vector retrieval and angular error calculation are repeated for all input polychromatic Stokes vector states, totaling 125 unique polychromatic polarization states. The same process is repeated for two more $c_{ref}$ values, for a total of three $c_{ref}$ values: $c_{ref}=1.44\pi$, $2.88\pi$, and $4.32\pi$. Figure \ref{fig:angular_error_box_plot} shows the statistics of both the simulated polychromatic Stokes vector retrieval and the experimental polychromatic Stokes vector retrieval.
\begin{figure}
    \centering
    \includegraphics[width=0.75\linewidth]{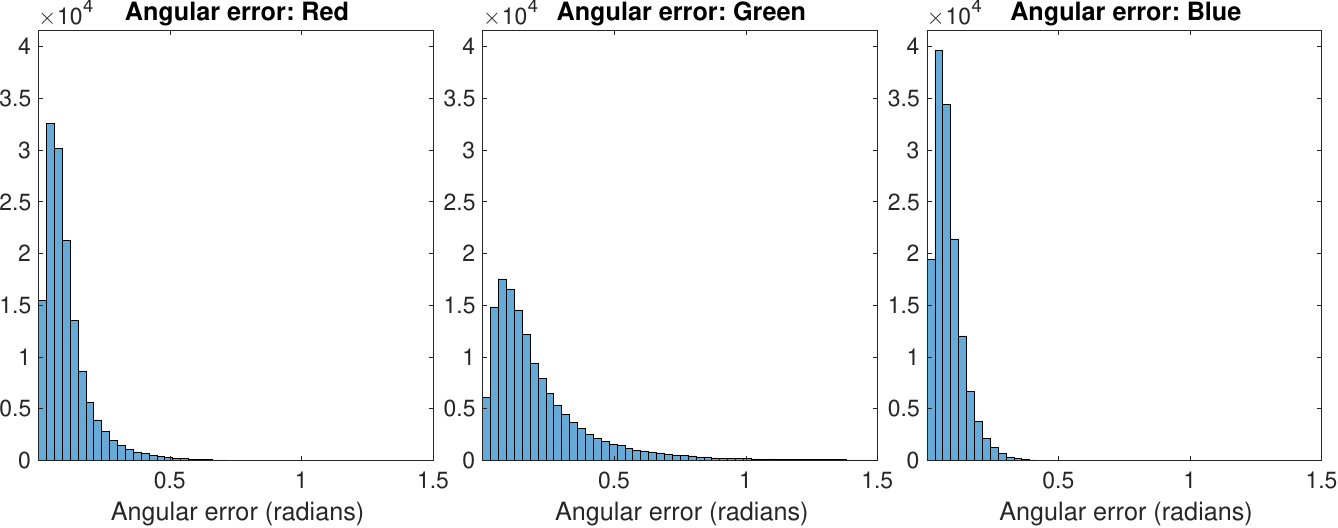}
    \caption{Histogram of the experimental data shown in Fig. \ref{fig:angular_error_box_plot} for $c_{ref}=2.88\pi$}
    \label{fig:angular_error_histo}
\end{figure}
The statistics of the angular error between the simulated and experimental polychromatic Stokes vector retrieval show good agreement. A major difference is that each input wavelength has less variations in angular error between the different $c_{ref}$ values. For $c_{ref}=1.44\pi$, the median angular error values for red, green, and blue respectively are 0.093, 0.194, and 0.076 radians. For $c_{ref}=2.88\pi$, the median angular error values for red, green, and blue respectively are 0.08, 0.156, and 0.068 radians. For $c_{ref}=4.32\pi$, the median angular error values for red, green, and blue respectively are 0.070, 0.149, 0.069 radians. A histogram of the experimental data shown in Fig. \ref{fig:angular_error_box_plot} for $c_{ref}=2.88\pi$ is shown in Fig. \ref{fig:angular_error_histo} for another view on the statistical behavior of angular error in the polychromatic Stokes vector retrieval. Those results are compared to the monochromatic Stokes vector retrieval, where the Stokes vector are retrieved for each input wavelength as individual, separate inputs as shown in Fig. \ref{fig:monochromatic_angular_error}. The monochromatic angular error is comparable to results obtained by Zimmerman and Brown \cite{zimmerman_star_2016} of 10 mrad (a precision that is achieved by averaging the recorded PSFs over several frames). 

\section{Discussion}
 \subsection*{Relationship between values of $c_{ref}$ and angular error}
Previous work\cite{spiecker_single-shot_2024} in modeling single-shot spectropolarimetry with an SEO explored how the PSF changes with the value of $c_{ref}$ and the relationship between value of $c_{ref}$ and angular error for a single point source. With spatially resolved spectropolarimetry, there are additional system parameters to consider, namely the lenslet pitch which corresponds to the distance between PSFs in an array of PSFs. Increasing $c_{ref}$ increases the PSF size, indicating a upper limit to the value of $c_{ref}$ for any given lenslet construction. Experimental results with the $c_{ref}$ values used seem to suggest that an increasing value of $c_{ref}$ leads to less angular error. However, once the PSFs grow in size, increasing the amount of cross talk, it is likely that angular error will begin to increase. Therefore, for a given system configuration, especially the lenslet construction, there exists an optimal value of $c_{ref}$ for a given input spectrum that may vary from one configuration to another. 

\subsection*{Spectral considerations}
The choice of input wavelengths of 635, 520, and 405 nm make for a good test case for single-shot spatially resolved spectropolarimetry. The wavelengths produced by the lasers used are in the visible spectrum, contain narrow linewidths (1-3 nanometers or less) and are evenly spaced on the spectrum. Additionally, commercially available RGB LEDs emit red, green, and blue light and such LEDs are used in displays which utilize polarization effects to produce a desired output. The light produced by such a display is a superposition of the input wavelengths, each potentially with different polarizations. Other examples of light sources where single-shot, spatially resolved spectropolarimetry would be applicable would be the light produced in second harmonic generation or Raman scattering. 

The model used for spectropolarimetry with an SEO assumes discrete wavelengths of infinitely narrow line widths spaced apart from each other, producing a well defined effective values of the dimensionless stress parameter $c$ for each input wavelength. Currently, the model propagates individual monochromatic fields through the system to obtain simulated results. A more realistic model would consist of spectral bins, with some specified spectral bin widths, and with some specified spacing between spectral bins. Future work in improving the model will understand better how more complex input spectra affects the accurate retrieval of the polychromatic Stokes vector.

\subsection*{Complex polychromatic polarization patterns}
\begin{figure}
    \centering
    \includegraphics[width=0.85\linewidth]{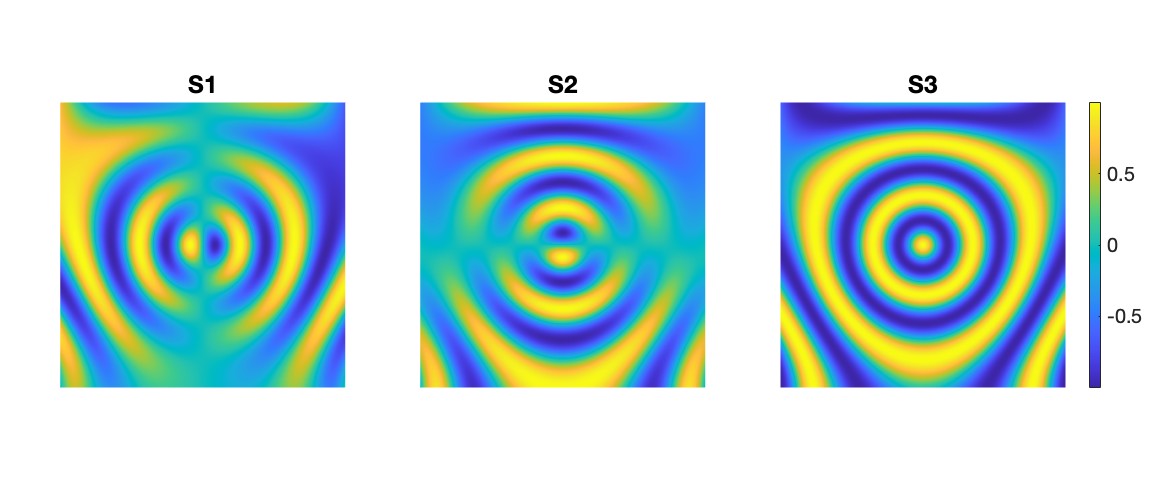}
    \caption{Simulated Stokes retrieval for a monochromatic red R-polarized input through an SEO. Note that the orientation of the fringes depends on where the stress points on the SEO are located, of importance are the relationship between S1, S2, and S3 patterns.}
    \label{fig:SEOred_simulated}
\end{figure}

\begin{figure}
    \centering
    \includegraphics[width=\linewidth]{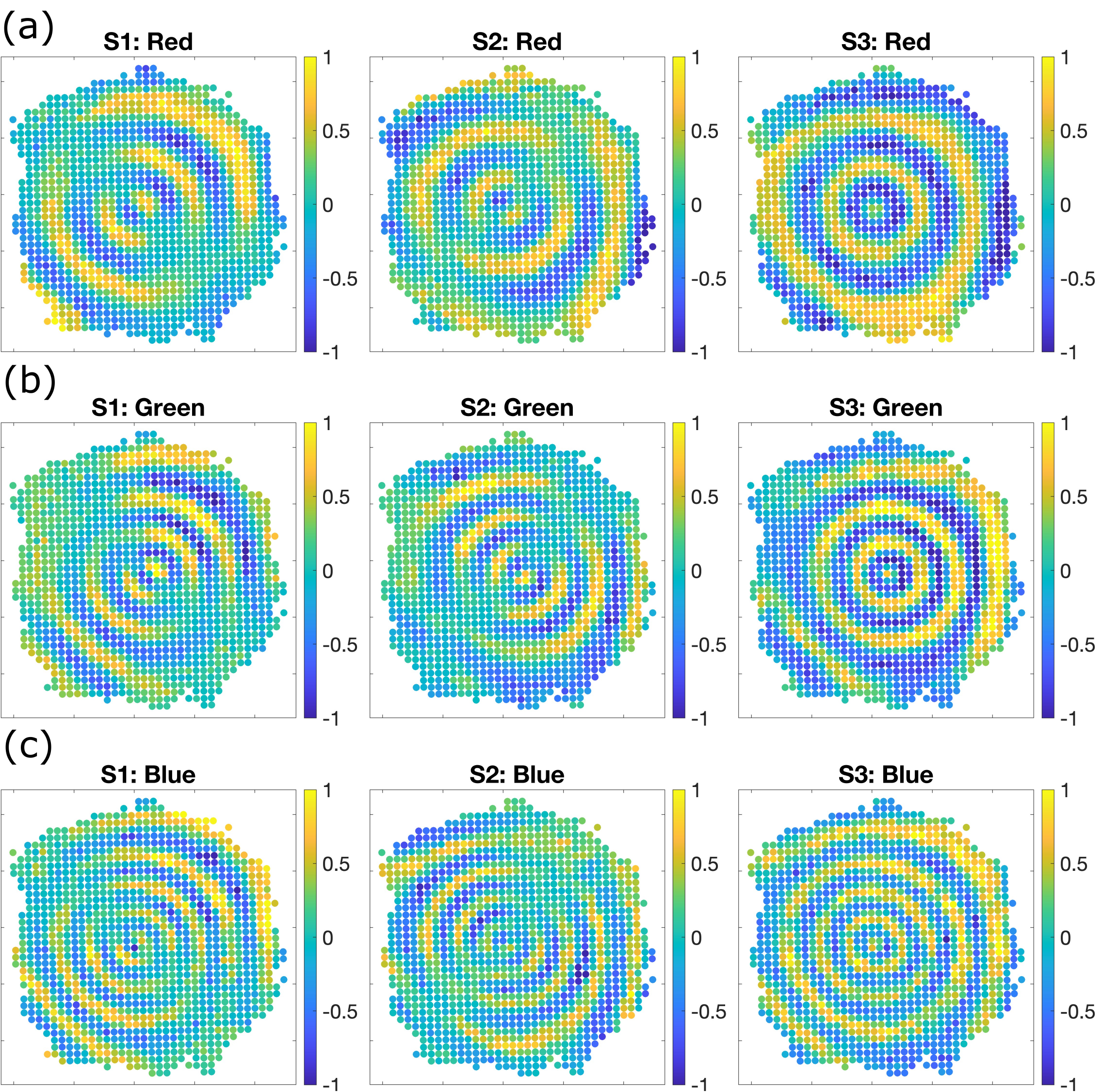}
    \caption{Spatially resolved monochromatic Stokes retrieval of a R-polarized input through an SEO. The wavelength dependence of the dimensionless stress parameter $c$ can be observed through the change in the fringe period. (a) Monochromatic red input. (b) Monochromatic green input. (c) Monochromatic blue input.}
    \label{fig:monochromatic_Stokes_retrieval_SEO}
\end{figure}

\begin{figure}
    \centering
    \includegraphics[width=\linewidth]{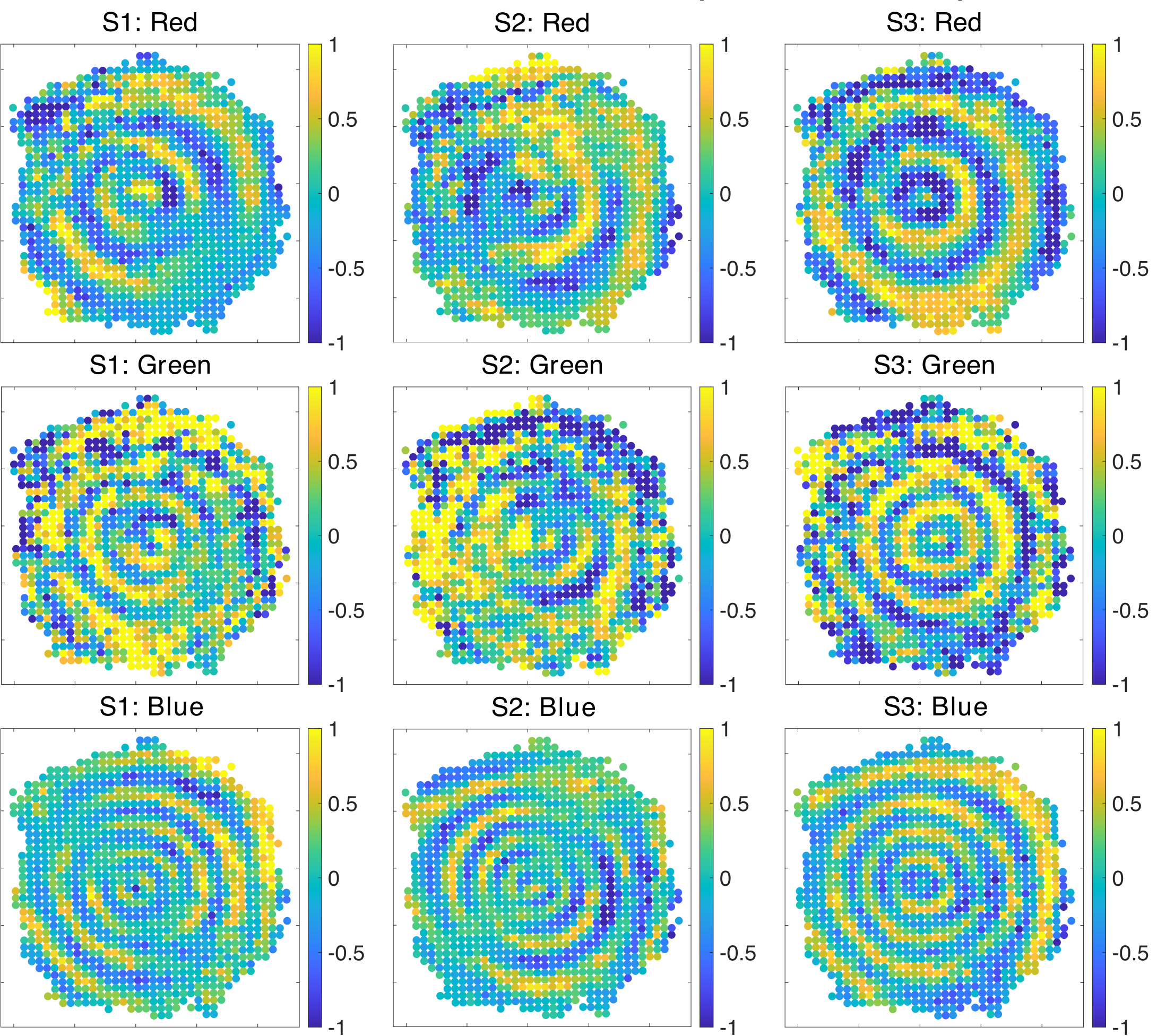}
    \caption{Polychromatic Stokes retrieval of a R-polarized input through an SEO. The polychromatic Stokes vector is retrieved by using the data shown in Fig. \ref{fig:monochromatic_Stokes_retrieval_SEO} for each input wavelength superimposed into a single polychromatic image.}
    \label{fig:polychromatic_Stokes_retrieval_SEO}
\end{figure}

An important assumption of the model used for single-shot spectropolarimetry with SEOs is that the input polarization pattern for any input wavelength is slowly varying. The polarization states produced by SEOs presents an interesting test case for the Stokes retrieval of complex polarization patterns. Figure \ref{fig:SEOred_simulated} shows the Stokes retrieval of a theoretical polarization state produced by sending a R-polarized input through an SEO as an input. Figure \ref{fig:monochromatic_Stokes_retrieval_SEO} shows a successful monochromatic Stokes vector retrieval when R-polarized input is sent through the same SEO for each input wavelength. Figure \ref{fig:polychromatic_Stokes_retrieval_SEO} shows the polychromatic Stokes vector retrieval using the same data collected for Fig. \ref{fig:monochromatic_Stokes_retrieval_SEO}. The results of the polychromatic Stokes vector retrieval highlights areas for further work in evaluating single-shot spectropolarimetry with SEOs for complex polarization patterns as inputs. 

\subsection*{Future direction}
In this work, the lenslet array successfully is used to produce an array of point sources in which single-shot, spatially resolved spectropolarimetry. The use of lenslet arrays was a deliberate choice because it leads to a future direction. Lenslet arrays are used in wavefront sensing in a Shack-Hartmann wavefront sensor. With a Shack-Hartmann wavefront sensor, each lenslet in the array samples the incident wavefront and produces a spot displacement corresponding to the incident wavefront gradient (assuming a slowly varying wavefront). For sufficiently small wavefront gradients incident on each lenslet, where the spot displacement induced by it are small enough, it is possible to perform single-shot, simultaneous spatially resolved spectropolarimetry and wavefront sensing with an SEO.  

\subsection*{Disclosures}
The authors declare that there are no financial interests, commercial affiliations, or other potential conflicts of interest that could have influenced the objectivity of this research or the writing of this paper.

\subsection* {Code, Data, and Materials Availability} 
Data underlying the results presented in this paper are not publicly available at this time but may be obtained from the authors upon reasonable request.

\subsection* {Acknowledgments}
We want to acknowledge previous work in modeling and polarimetry with SEOs by the following former members of, and collaborators with, the Brown lab: Alexis Vogt, Amber Beckley, Roshita Ramkhalawon, Ashan Ariyawansa, Brandon Zimmermann, and Miguel Alonso. Likewise, we want to acknowledge people who have provided valuable advice and discussion: Brian Kruschwitz and Katelynn Bauer. Furthermore, we acknowledge financial support from: the Horton Fellowship from the Laboratory for Laser Energetics, the GAANN fellowship from the US Department of Education (P200A210035-322 23), the Institute of Optics at University of Rochester, and SPIE.

%%%%% References %%%%%

\bibliography{references}

\begin{thebibliography}{10}

\bibitem{sabatke_snapshot_2002}
D.~S. Sabatke, A.~M. Locke, E.~L. Dereniak, {\em et~al.}, ``Snapshot imaging
  spectropolarimeter,'' {\em Optical Engineering} {\bf 41}, 1048--1054  (2002).
\newblock doi:10.1117/1.1467934.

\bibitem{todorov_spectrophotopolarimeter_1992}
T.~Todorov and L.~Nikolova, ``Spectrophotopolarimeter: fast simultaneous
  real-time measurement of light parameters,'' {\em Optics Letters} {\bf 17},
  358--359  (1992).
\newblock doi:10.1364/OL.17.000358.

\bibitem{oka_spectroscopic_1999}
K.~Oka and T.~Kato, ``Spectroscopic polarimetry with a channeled spectrum,''
  {\em Optics Letters} {\bf 24}, 1475--1477  (1999).
\newblock doi:10.1364/OL.24.001475.

\bibitem{knitter_spectrally_2011}
S.~Knitter, T.~Hellwig, M.~Kues, {\em et~al.}, ``Spectrally resolving
  single-shot polarimeter,'' {\em Optics Letters} {\bf 36}, 3048  (2011).
\newblock doi:10.1364/OL.36.003048.

\bibitem{tyo_review_2006}
J.~S. Tyo, D.~L. Goldstein, D.~B. Chenault, {\em et~al.}, ``Review of passive
  imaging polarimetry for remote sensing applications,'' {\em Applied Optics}
  {\bf 45}, 5453--5469  (2006).
\newblock doi:10.1364/AO.45.005453.

\bibitem{rubin_matrix_2019}
N.~A. Rubin, G.~D’Aversa, P.~Chevalier, {\em et~al.}, ``Matrix {Fourier}
  optics enables a compact full-{Stokes} polarization camera,'' {\em Science}
  {\bf 365}, eaax1839  (2019).
\newblock doi:10.1126/science.aax1839.

\bibitem{rubin_imaging_2022}
N.~A. Rubin, P.~Chevalier, M.~Juhl, {\em et~al.}, ``Imaging polarimetry through
  metasurface polarization gratings,'' {\em Optics Express} {\bf 30},
  9389--9412  (2022).
\newblock doi:10.1364/OE.450941.

\bibitem{li_metasurface-enabled_2025}
L.~W. Li, P.~H.~H. Oakley, R.~N. Schindhelm, {\em et~al.},
  ``Metasurface-{Enabled} {Astronomical} {Polarimetry},''  (2025).
\newblock doi:10.48550/arXiv.2506.06245.

\bibitem{zimmerman_star_2016}
B.~G. Zimmerman and T.~G. Brown, ``Star test image-sampling polarimeter,'' {\em
  Optics Express} {\bf 24}, 23154--23161  (2016).
\newblock doi:10.1364/OE.24.023154.

\bibitem{sivankutty_single-shot_2016}
S.~Sivankutty, E.~R. Andresen, G.~Bouwmans, {\em et~al.}, ``Single-shot
  polarimetry imaging of multicore fiber,'' {\em Optics Letters} {\bf 41},
  2105--2108  (2016).
\newblock doi:10.1364/OL.41.002105.

\bibitem{yiannopoulos_general_1999}
A.~Yiannopoulos, ``A {General} {Formulation} of {Stress} {Distribution} in
  {Cylinders} {Subjected} to {Non}-{Uniform} {External} {Pressure},'' {\em
  Journal of Elasticity} {\bf 56}, 181--198  (1999).
\newblock doi:10.1023/A:1007667200738.

\bibitem{brown_stress_2013}
T.~G. Brown and A.~M. Beckley, ``Stress engineering and the applications of
  inhomogeneously polarized optical fields,'' {\em Frontiers of
  Optoelectronics} {\bf 6}, 89--96  (2013).
\newblock doi:10.1007/s12200-012-0307-5.

\bibitem{ramkhalawon_star_2012}
R.~Ramkhalawon, A.~M. Beckley, and T.~G. Brown, ``Star test polarimetry using
  stress-engineered optical elements,'' in {\em Three-{Dimensional} and
  {Multidimensional} {Microscopy}: {Image} {Acquisition} and {Processing}
  {XIX}},   {\bf 8227}, 125--132, SPIE  (2012).
\newblock doi:10.1117/12.908472.

\bibitem{ramkhalawon_imaging_2013}
R.~D. Ramkhalawon, T.~G. Brown, and M.~A. Alonso, ``Imaging the polarization of
  a light field,'' {\em Optics Express} {\bf 21}, 4106--4115  (2013).
\newblock doi:10.1364/OE.21.004106.

\bibitem{spiecker_single-shot_2024}
D.~Spiecker and T.~G. Brown, ``Single-shot spectropolarimetry with
  stress-engineered optics,'' in {\em Optical {Modeling} and {Performance}
  {Predictions} {XIV}},   {\bf 13129}, 48--56, SPIE  (2024).
\newblock doi:10.1117/12.3044380.

\bibitem{spiecker_stress_2025}
D.~Spiecker and E.~Herger, ``Stress {Engineered} {Optics}: {Optomechanical}
  {Design} and {Performance},''  (2025).
\newblock doi:10.1364/opticaopen.29640998.v1.

\end{thebibliography}
\bibliographystyle{spiejour}   % makes bibtex use spiejour.bst

\listoffigures
\listoftables

\end{spacing}
\end{document}